\documentclass[useAMS,usenatbib]{mn2e}
\usepackage{txfonts}
\usepackage{rotating}
\usepackage{textcomp}
\usepackage{amssymb}
\usepackage{graphicx}
\usepackage{txfonts}
\usepackage{subfig}
\usepackage{float}

\usepackage{url}
\usepackage{natbib}
\usepackage{booktabs}
\usepackage[normalem]{ulem}
\usepackage{colortbl, xcolor}
\usepackage{multirow}
\usepackage{threeparttable}

\def\gray{\rowcolor[gray]{0.9}}

\def\kkms{K~km~s$^{-1}$}
\def\kms{km~s$^{-1}$}
\def\chisq{$\chi^2$ }
\def\msunpcthree{$\rm M_{\odot}\,pc^{-3}$ }
\def\msunpctwo{$\rm M_{\odot}\,pc^{-2}$ }
\def\microm{\mbox{$\rm\mu m$ }}
\def\msunyr{\mbox{$\rm M_{\odot}\,yr^{-1}$}}
\def\xco{$X_{\rm CO}\,$}
\def\SFR{$\Sigma_{\rm SFR}\,$}

\def\cof{\mbox{$^{12}$CO~(1--0)}}
\def\cos{\mbox{$^{12}$CO~(2--1)}}
\def\cot{\mbox{$^{12}$CO~(3--2)}}
\def\tcos{\mbox{$^{13}$CO~(2--1)}}
\def\tcot{\mbox{$^{13}$CO~(3--2)}}

\title[Physical Condition of Molecular Gas at the Centre of NGC\,1097]{Physical Condition of Molecular Gas at the Centre of NGC\,1097}
\author[N. Pi\~nol-Ferrer et al.]
{N. Pi\~nol-Ferrer$^{1,2}$, K. Fathi$^{1,2}$, A. Lundgren$^3$, G. van de Ven$^{4}$\\
$^1$Stockholm Observatory, Department of Astronomy, Stockholm University, AlbaNova Center, 106 91 Stockholm, Sweden\\
$^2$Oskar Klein Centre for Cosmoparticle Physics, Stockholm University, 106 91 Stockholm, Sweden\\
$^3$European Southern Observatory, Casilla 19001, Santiago 19, Chile\\
$^4$Max Planck Institute for Astronomy, Knigstuhl 17, 69117 Heidelberg, Germany
}
\date{Accepted for publication, 2011 January 25}

\pagerange{\pageref{firstpage}--\pageref{lastpage}} \pubyear{2010}

\begin{document}

\label{firstpage}

\maketitle

\begin{abstract}
We have used the \xco\ conversion factor, Local Thermal Equilibrium and Large Velocity Gradient approximation to parametrize the cold and warm phase of the interstellar medium from five different low transitions of the CO molecule in the central 21\arcsec\ (kpc) region of NGC\,1097. We have applied a one-component model and derived a typical kinetic temperature of about 33~K, a molecular Hydrogen density of $4.9\times 10^{3}$ \msunpcthree and a CO column density of $1.2 \times 10^{-2}$ \msunpctwo. A two-component model results in 85\% cold-to-total gas fraction in the presence of a 90 K warm counterpart. Furthermore, we ``resolve" the spatially unresolved single dish observations by selecting velocity channels that in an interferometric velocity map correspond to specific regions. We have selected five such regions and found that the physical properties in these regions are comparable to those derived from the full line profile. This implies that the central kpc of NGC\,1097 is rather homogeneous in nature, and, although the regions are not uniquely located within the ring, the star formation along the ring is homogeneously distributed (in agreement with recent Herschel observations). We have further revised the mass inflow rate onto the Supermassive Black Hole in this prototype LINER/Sy1 galaxy and found that, accounting for the total interstellar medium and applying a careful contribution of the disc thickness and corresponding stability criterion, increases the previous estimations by a factor 10. Finally we have calculated the \xco\ conversion factor for the centre of NGC\,1097 using an independent estimation of the surface density to the CO emission, and obtained $X_{\rm CO}=(2.8\pm0.5)\times 10^{20} \, \rm cm^{-2}\, (\rm K\, km\, s^{-1})^{-1}$ at radius 10.5\arcsec\ and $X_{\rm CO}=(5.0\pm0.5)\times 10^{20} \, \rm cm^{-2}\, (\rm K\, km\, s^{-1})^{-1}$ at radius 7.5\arcsec. With the approach and analysis described in this paper we have demonstrated that important physical properties can be derived to a resolution beyond the single dish resolution element, however, caution is necessary for interpreting the results.
\end{abstract}


\section{Introduction} 

The connection between star formation in centres of galaxies and the onset of nuclear activity involve processes still not completely understood. In many cases, the majority of star formation in galaxies is located in circumnuclear rings, and the most extended mechanism for fuelling them and generate densities high enough to form stars, is believed to be non-axisymmetric perturbations, such as bars or tidal interactions. In the presence of a bar, over-densities of gas may occur in resonances, like Inner Lindblad Resonance (ILR) for a inner ring (e.g. \citealt{CombesGerin1985}, \citealt{Athanassoula1992}, \citealt{Wada1994}). Once the gas is accumulated in the ring, there are two main hypothetical scenarios for the formation of the stars in the ring: (1) star formation produced by gravitational instabilities in different giant clouds placed randomly along the ring \citep{Elmegreen1994}, and (2) quasi-instantaneous star formation in two regions with over densities, usually the connection points of the dust lanes of the bar with the ring, followed by migration of the star forming knots along the rotation velocity of the disc, producing an age gradient along the ring \citep{Boker2008}.

The study of star formation locations is typically done observing H$\rm \alpha$ or radio continuum emission. In addition, it is possible to use direct measurements of gas column density ($\Sigma_{\rm gas}=\Sigma_{\rm HI}+\Sigma_{\rm H_2}$), to which the star formation rate per unit area \SFR is empirically related by the Schmidt law (\citealt{Schmidt1959}, \citealt{Kennicutt1989}). One aim of this paper is to measure the molecular gas column density in order to characterize the star formation along the 1 kpc ring of the nearby prototype LINER/Sy1 barred galaxy NGC\,1097, estimated through CO observations.

The fuelling of nuclear and/or circumnuclear activity in barred galaxies is driven by the bar itself due to loss of angular momentum along the dust lanes (e.g. \citealt{Athanassoula1992}), however, the fate of the inflow mass after passing the ILR is still unsolved. Several mechanisms have been inferred to explain the fate of the inflowing gas. One proposed mechanism is the coupling of nested bars inside the ILR of primary bars, transporting gas and stars towards the centre (e.g. \citealt{Shlosman1989}, \citealt{EnglmaierShlosman2004}). This mechanism is not completely established since there are other studies which show that nested bars are not allow to carry material to the inner few parsec of galaxies \citep{Maciejewski2002}. Another proposed mechanism is inflow along nuclear spiral arms, which are connected with the dust lanes of the primary bar (e.g. \citealt{EnglmaierShlosman2000}, \citealt{Maciejewski2002}, \citealt{Maciejewski2004b}). These arms have a very low density contrast compared with the ambient inter-arm regions, only about the 10\% \citep{EnglmaierShlosman2000}, and present a wide variety of observed morphologies. Morphologically, nuclear spirals are divided into two main categories, chaotic spirals, and symmetric or grand design spirals, for which different formation mechanisms have been proposed (e.g., \citealt{Elmegreen1998};  \citealt{EnglmaierShlosman2000}; \citealt{Maciejewski2002}). In the presence of a central super massive black hole (SMBH), spiral shocks can extend all the way to its vicinity and generate gas inflow compatible to the accretion rates observed in local active galactic nuclei.  (e.g. morphologically by \citealt{Martini2003} and kinematically by \citealt{Fathi2006}).

NGC\,1097 is an excellent candidate to study the inflow processes driven by the bar, however, the lack of direct measurements of the physical conditions of the interstellar gas in the circumnuclear regions has complicated matters. Optical \citep{vandeVenFathi2010}; Near infra-red  (\citealt{Davies2009} and \citealt{Kotilainen2000}), radio continuum \citep{Beck2005}, CO transitions with radio interferometers and single dish (\citealt{Kohno2003}, \citealt{petitpas2003} \citealt{Hsieh2008}) and far infra-red observations (\citealt{HerschelI}, \citealt{HerschelII}) have been performed for its 1 kpc starburst ring. Moreover the still unclear processes involved in the star formation in the ring, the estimation of the mass accretion rate toward its $(1.2 \pm 0.2)\times 10^8 \, M_{\odot}$ SMBH \citep{LewisEracleous2006} is to date inconclusive. As a second goal, we shed light on the total gas mass in the region inside the ring, and apply several methods to better understand the physical state of the cold+warm ($<150$ K) gas in the circumnuclear kpc of NGC\,1097.

We use archival data combined with our own single dish observations of the molecular gas in the circumnuclear regions using the Atacama Pathfinder EXperiment (APEX) telescope. The data are used with two different line ratio modelling methods to obtain a new estimation of the physical properties of the gas and the accretion mass rate towards the SMBH to compare with the previous estimations. Moreover, we derive the molecular gas physical conditions along regions with comparable line of sight velocities, and in this way override the limited spatial resolution of the single dish observations. 
In section~\ref{sec:obs}, we present the observations and data reduction. In section~\ref{sec:analysis}, we explain all the analysis done for estimating the line ratios, as for study the physical properties using the \xco\ conversion factor, a local thermal equilibrium and a large velocity gradient approximation for the inner kpc of NGC\,1097. In section~\ref{sec:gasring}, we explain the analysis made within the ring and discuss the result obtained there, in section~\ref{sec:mir} we discuss the mass inflow rate towards the inner parsecs, and in section~\ref{sec:xco} we present a study of the real \xco\ conversion factor value for this galaxy. Finally, we conclude in section~\ref{sec:conclusions}.

\section{Observations and Data Reduction}
\label{sec:obs}
\begin{figure*}
   \centering
\includegraphics[width=0.98\textwidth,trim= 0mm 10mm 10mm 0mm]{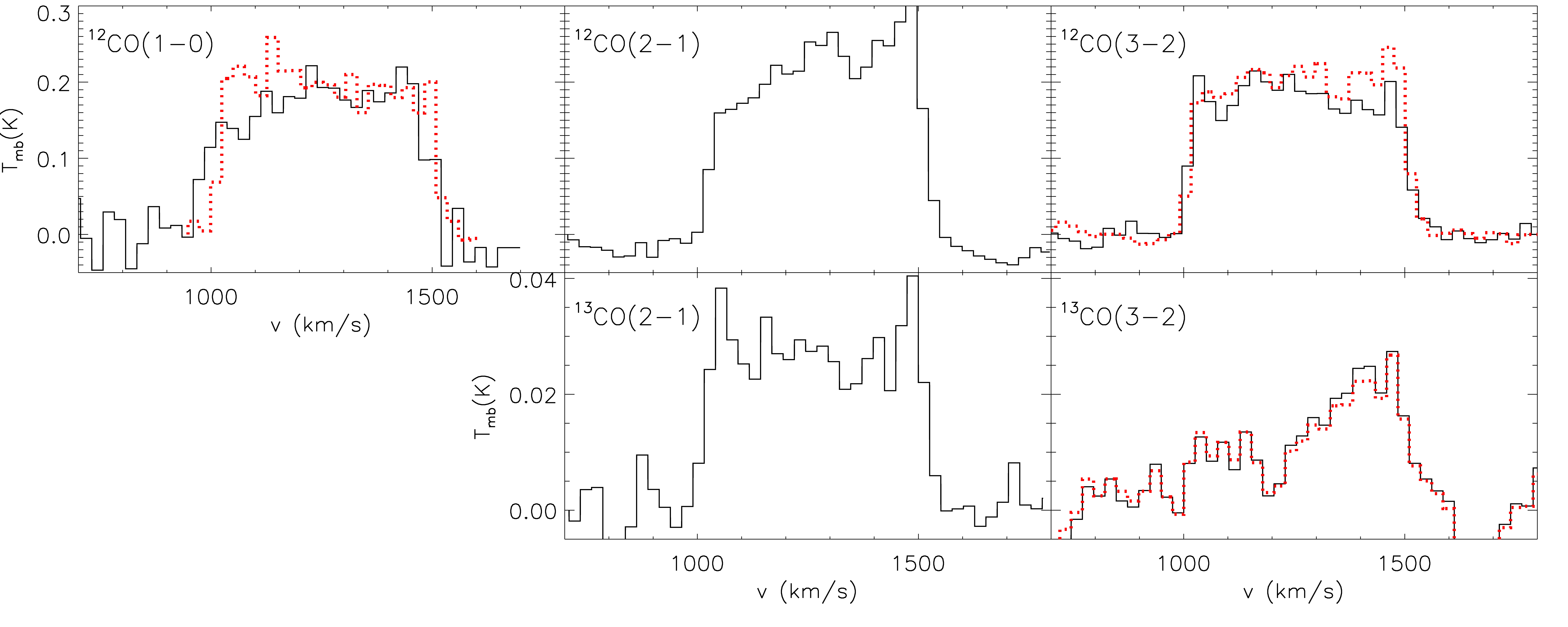}
\caption{The different observations with corresponding transition indicated in each panel at 21\arcsec\ resolution (black curves) as described in section~\ref{sec:obs}. The original 15\arcsec\ \cof, 18\arcsec\  \cot\ and mapped \tcot\ lines are over-plotted with red dotted curves.}
\label{fig:lines}
\end{figure*}

On the 27th of June of 2008, we observed \cot\ and \tcot\ in one pointing at the centre of NGC\,1097 with the Atacama Pathfinder Experiment (APEX) on Llano de Chajnantor, Chile, using the APEX-2 receiver. The Half-Power Beam Width (HPBW) is 18\arcsec\ at 345.8 GHz and 18.8\arcsec\ at 330 GHz. In addition, we mapped \cot\ along the primary bar of NGC\,1097 on the 7th of July of 2009, using the same set up. We mapped the bar with an array of 9$\times$3 points. We scaled the antenna temperature adopting the value for $\eta_{MB}$ presented in the APEX web page of 0.73. During the reduction of the spectra, we subtracted a first order baseline from the profiles. Lines are shown in Fig.~\ref{fig:lines}.

Our data have been combined with published single dish and interferometric data from the centre of this galaxy: \cof\ from \citet{Kohno2003} with a HPBW of 15\arcsec\ observed with the Nobeyama Radio Observatory, \cos\ and \tcos\ from \citet{petitpas2003} with a HPBW of 21\arcsec\ observed with James Clerk Maxwell Telescope, and \cof\ observed with Nobeyama Millimetre Array \citep{Kohno2003}. All the auxiliary data were convolved to the common beam size of 21\arcsec, and we discuss \cof\ case separately in the following subsection.
For \tcot\ transition we used to estimate the ratio between \cot\ and \tcot, both with a beam size of 18\arcsec, we assume that the ratio changes only marginally between 18\arcsec\ and 21\arcsec\ beam sizes. 
We further observed the \tcot\ in additional nine positions around the central 18.8\arcsec, with the APEX telescope, 8th of October of 2010 and confirm the validity of the bottom right panel of Fig.~\ref{fig:lines} as representative of a 21\arcsec\ observations.

\subsection{\cof\ }
We combined the \cof\ single dish line with the \cof\ interferometric data from \citet{Kohno2003} to estimate the missing flux. The interferometric data were thus convolved to the resolution of the single dish observations and a spectral line from the central beam of the galaxy was extracted, resulting in a missing flux of 7.7\%. Given the relatively marginal missing flux ($<10\%$), emission lines corresponding to 21\arcsec\ beam could be obtained simply by co-addition of the interferometric data.

Furthermore, we improve our estimation of the line using the average of two different methods to recover the 21\arcsec\ line. Method~1: adding 7.7\% to the co-added interferometric data, and Method~2: by calculating the missing flux for every channel of the line with 15\arcsec\ beam, and adding the corresponding missing flux (per channel) to the corresponding channel of the generated 21\arcsec\ line. Following these tests, for our analysis, we use the average of these two cases, for which a flux error can be obtained as the mean of the standard deviations of every channel.

\section{The unresolved central kpc}
\label{sec:analysis}

\subsection{Line-ratio estimation}
All lines were binned to the velocity sampling of 26 \kms\ to match the largest sampling available to us. We neglected channels with a signal-to-noise ($S/N$) smaller than 3, and calculated the ratio of the different lines by averaging the ratios between temperatures of every channel. The corresponding errors were then the standard deviation of this average. We also estimated the ratios using the velocity-integrated intensity of every line, and found that they agree very well with the previous measurement. The results are shown in the tables \ref{tab:intensities} and \ref{tab:ratios}.

\begin{table*}
\caption{Spectral line parameters and specifications.}
 \label{tab:intensities}
  \centering\tabcolsep=20pt
    \begin{threeparttable}
       \begin{tabular}{c|ccc}
       \toprule
       \multirow{2}{*}{\textbf{Line}} & \textbf{Beam size (HPBW)} & \textbf{Intensity} $\mathbf{\int T_{mb}dV}$ & \textbf{Reference}  \\
          & \arcsec\ &  \kkms\ & \\
           \midrule
      \gray $^{12}$CO (1-0) & 21/15 & $89 \pm 9$/ $96\pm2.8$ \tnote{a}& \citet{Kohno2003} \\
       $^{12}$CO (2-1) & 21 & $110 \pm 7$ & \citet{petitpas2003} $(121 \pm 1$ \kkms)\tnote{a}\\
          \gray $^{12}$CO (3-2) & 21/18& $91 \pm 6$/$102 \pm 4$ & APEX data\tnote{b} \\
       $^{13}$CO (2-1) & 21 & $ 14 \pm 2 $& \citet{petitpas2003} $(14.6 \pm 0.6$ \kkms)\tnote{a}\\
      \gray $^{13}$CO (3-2) & 18/21 & $7 \pm 3$/$6\pm 3$ & APEX data\\
       \bottomrule
     \end{tabular}
  \begin{tablenotes}
  \item[a]Previous estimation of the intensities by corresponding author.
  \item[b]\citet{petitpas2003} derived $121.7 \pm 0.5$ \kkms for 21\arcsec\ beam.
  \end{tablenotes}
\end{threeparttable}
\end{table*}

The errors which contribute to the line measurements are: ($i$) the rms, ($ii$) the error due to bad pointing, ($iii$) the missing flux of \cof\ and ($iv$) the uncertainty of the main beam efficiency $\eta_{MB}$ and that of the Main Beam size $\theta_{MB}$. Our data reduction and re-sampling method provides a good estimation of the rms and missing flux errors, and the errors due to beam size and efficiency, together, are estimated to be of the order of 10\% (at most 10\% for the main beam efficiency and 0.05\% for the main beam size). The error due to bad pointing was calculated using the interferometric data for \cof\ and the error pointing of each telescope (2\arcsec\ rms for APEX and 3\arcsec\ rms for the data from \citet{petitpas2003}). We studied the difference of the lines produced by varying the centre within the error in the pointing, and find that this effect corresponds to additional 11\% for 2\arcsec\ and 15\% for 3\arcsec\ rms.

\begin{table*}
  \center
\caption{Observed line ratios with average errors.}
\label{tab:ratios}
   \centering\tabcolsep=20pt
   \begin{tabular}{c | c c c c }
   \toprule
   & $\mathbf{\frac{^{12}CO (1-0)}{^{12}CO (2-1)}}$  & $\mathbf{\frac{^{12}CO (3-2)}{^{12}CO (2-1)}}$ & $\mathbf{\frac{^{12}CO (2-1)}{^{13}CO (2-1)}}$&$\mathbf{\frac{^{12}CO (3-2)}{^{13}CO (3-2)}}$ \\
   \midrule
  Average temperature ratios &$0.77 \pm 0.15$ & $0.8 \pm 0.2$  & $7.9 \pm 1.6$ & $18 \pm 10$ \\
  Velocity-integrated line ratios &$0.8\pm 0.4$ & $0.8 \pm 0.3 $  & $8 \pm 4$ & $14 \pm 10$ \\
   \bottomrule
\end{tabular}
\end{table*}

\subsection{Local Thermal Equilibrium and \xco\ Conversion Factor Analysis}
\label{subsec:xco}
With the estimated intensities and ratios of the lines, we used two different methods to estimate the molecular gas physical properties in the inner kpc (21\arcsec) of NGC\,1097, (i) the \xco\ conversion factor and (ii) the Local Thermal Equilibrium (LTE) approximation. Although, none of these methods are fully realistic for our unresolved data, they are instructive to compare with the estimations using a Large Velocity Gradient (LVG) approximation.

Using the \xco conversion factor and the line \cof\ we estimated the H$_2$ column density of the inner 21\arcsec. We used a value for \xco\ of $3 \times 10^{20}$ $\rm cm^{-2}(\rm K \,km \,s^{-1})^{-1}$ (\citealt{Scoville1987} and \citealt{Solomon1987}, also used by \citealt{Hsieh2008}) for an illustrative estimation of the molecular gas density. With our estimation of the integrated intensity of the line, we obtained a column density of $(2.7 \pm 0.3) \times 10^{22} \,\rm cm^{-2}$. This value is of the same order as the estimation made by \citealt{Hsieh2008} at the radius of the starburst ring in this galaxy.
However, using the conversion factor obtained from outside the Galactic plane \xco$=1.8 \times 10^{20}$ $\rm cm^{-2}(\rm K \,km \,s^{-1})^{-1}$ \citep{Dame2001} implies a column density of $(1.6 \pm 0.2) \times 10^{22} \,\rm cm^{-2}$, and \xco$=0.4 \times 10^{20}$ $\rm cm^{-2}(\rm K \,km \,s^{-1})^{-1}$ \citep{Martin2010} gives a column density of $(3.6 \pm 0.4) \times 10^{21} \,\rm cm^{-2}$.

\begin{figure}
   \centering
\includegraphics[width=0.49\textwidth]{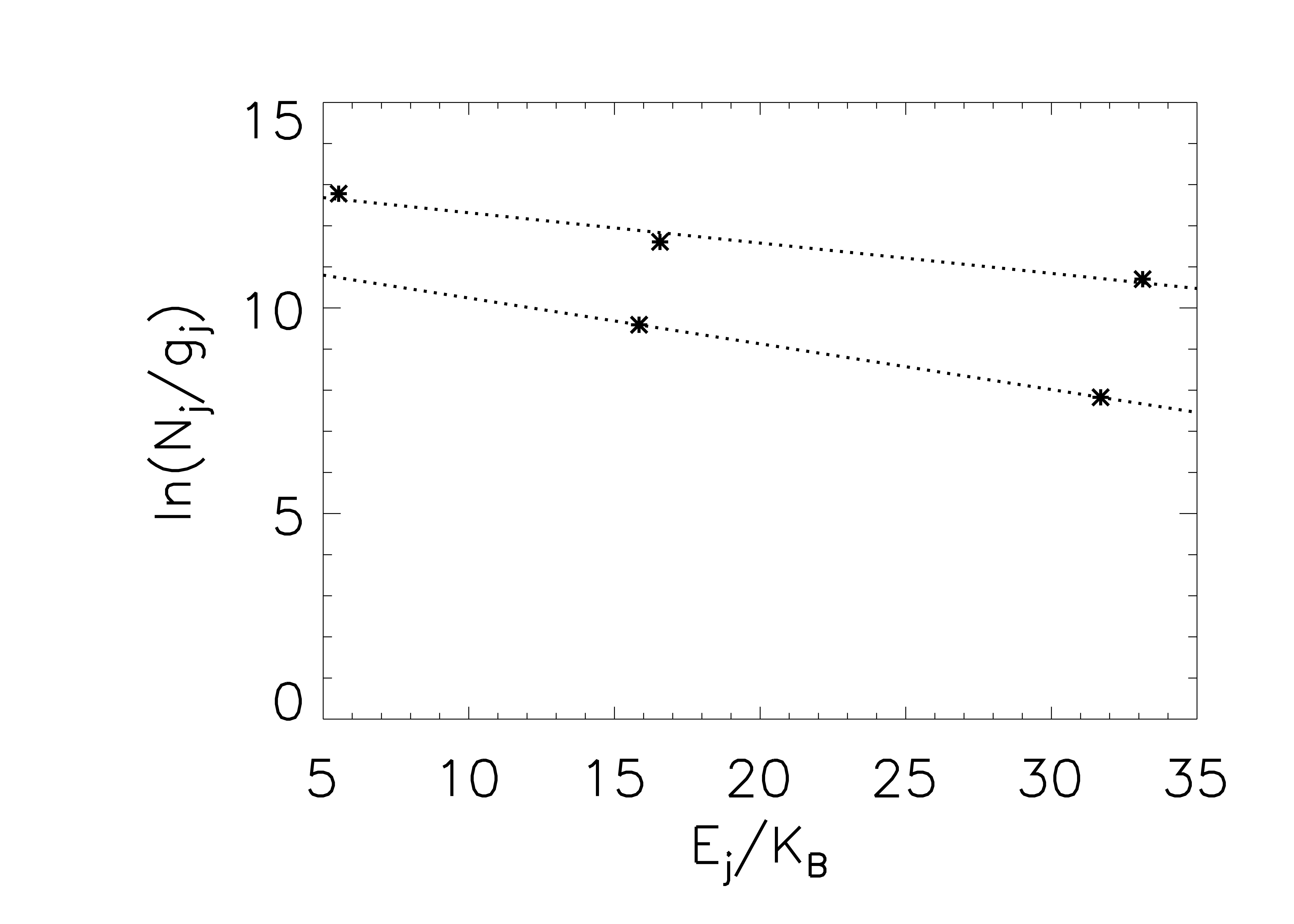}
\caption{Population diagram assuming LTE, where $N_j$  is the column density of the molecules with the energy state $j\,$  ($E_j$) and $g_j$  the statistical weight for $E_j$. The slope of these lines gives the kinetic temperature of the clouds. The upper line represents the $^{12}$CO cloud and the lower line, $^{13}$CO.}
\label{fig:lte}
\end{figure}

We also analysed the emission assuming LTE, making use of a ``Population Diagram" \citep{lte_GoldsmithLanger99}, shown in Figure \ref{fig:lte}. The LTE approximation gives a linear relation between ln$(N_j/g_j)$ and $E_j/k_B$ for a cloud, where $N_j$ is the column density of the molecules with the energy state $j\,$  ($E_j$) and $g_j$ the statistical weight for $E_j$. The slope of this linear relation, is inversely proportional to the kinetic temperature characteristic of the cloud. We estimated a typical kinetic temperature for $^{12}$CO and $^{13}$CO of about 10 K. In the case of an optically thick medium, this approximation only yields a lower limit for the column density. We derive a value that is indeed too low, which is due to the fact that in our LTE approximation we do not account for optical thickness. 

\subsection{Large Velocity Gradient Analysis}
\label{subsec:LVG}
The interferometric images of \cof\ \citep{Kohno2003} and \cos\ \citep{Hsieh2008}, show that most of CO emission comes from the very centre and the starburst nuclear ring of the galaxy. In reality, the connection points of the ring with the dust lanes of the primary bar present a clear increase of the emission. In the following subsections, we analyse the molecular gas physical properties of these regions performing a radiative transfer analysis of the single dish data described in section~\ref{sec:obs}.

To improve the estimation of the parameters presented in the previous section, we modelled the ratios of the observed lines using a non-Local Thermal Equilibrium (Non-LTE) radiative transfer code, \textsc{Radex} \citep{radex}, available at \texttt{http://www.strw.leidenuniv.nl/$\sim$moldata/radex.html}. \textsc{Radex} assumes the Large Velocity Gradient (LVG) approximation, and gives the expected line intensity of a molecular cloud with a characteristic kinetic temperature $T_K$, molecular Hydrogen density $n$(H$_2$), $^{12}$CO column density $\Sigma$($^{12}$CO), and line width $v$($^{12}$CO).

We studied these properties using a $\chi^2$ minimization comparing the observed line ratios and the ratios generated by \textsc{Radex}. We built a three-dimensional matrix made of different $\chi^2$ values created by changing the three principal input properties in the radiative transfer code, and varied the inputs within the ranges: 10--150 K (40 steps) for the kinetic temperature, 10--10$^6$ cm$^{-3}$ (50 steps) for the molecular Hydrogen density, and 10$^{15}$--10$^{18}$ cm$^{-2}$ (30 steps) for the $^{12}$CO column density. We fixed the internal $^{12}$CO line width to 1 \kms\ and the abundance ratio [$^{12}$CO]/[$^{13}$CO] to 40 \citep[c.f., $40 \pm 10$ found for the centre of active galaxies; ][]{Henkel1993, Mauersberger&HenkelReview1993, Henkel98, Bayet2004, Israel2009}. We further confirm that changing the abundance ratios between 30 and 50 introduces only marginal changes in the final results.

After fixing the abundance ratio and the line width, there are only three free parameters (namely temperature, density and column density) in the scenario which depict the four observed ratios that we are analysing here. However, there are still several possible solutions among which we select the best alternative using the combination of two criteria: the optical depth should be in agreement with previous findings \citep[e.g., $<10$ as previously found by ][]{petitpas2003} and minimizing the combination of the individual $\chi^2$ for every ratio. The errors were estimated using a bootstrap method and by looking at the fluctuation of each parameter around the $\chi^2$ minimum (see Fig. \ref{fig:contmin_1c_c40_v1}, where the red curves mark the 1$\sigma$ $\chi^2$ confidence level contours).

\subsection{The Inner 21\arcsec: One-component Model}
\label{subsec:Gasgloblal}
The circumnuclear 21\arcsec\ diameter hosts the active nucleus as well as the inner starburst ring with radius $\approx10$\arcsec. Our single dish observations are affected by beam-dilution effect in the measured line strength, since the size of the beam is larger than the individual clouds, and the clumpiness of the clouds and the large beam size produce an unknown filling factor. However, using line ratios with an equal beam size, instead of the full line profile, and assuming that the emission of every transition comes from the same clouds, these problems are avoided (see also \citealt{petitpas2003}, \citealt{Israel2009}).

\begin{figure*}
   \centering
\includegraphics[width=0.98\textwidth]{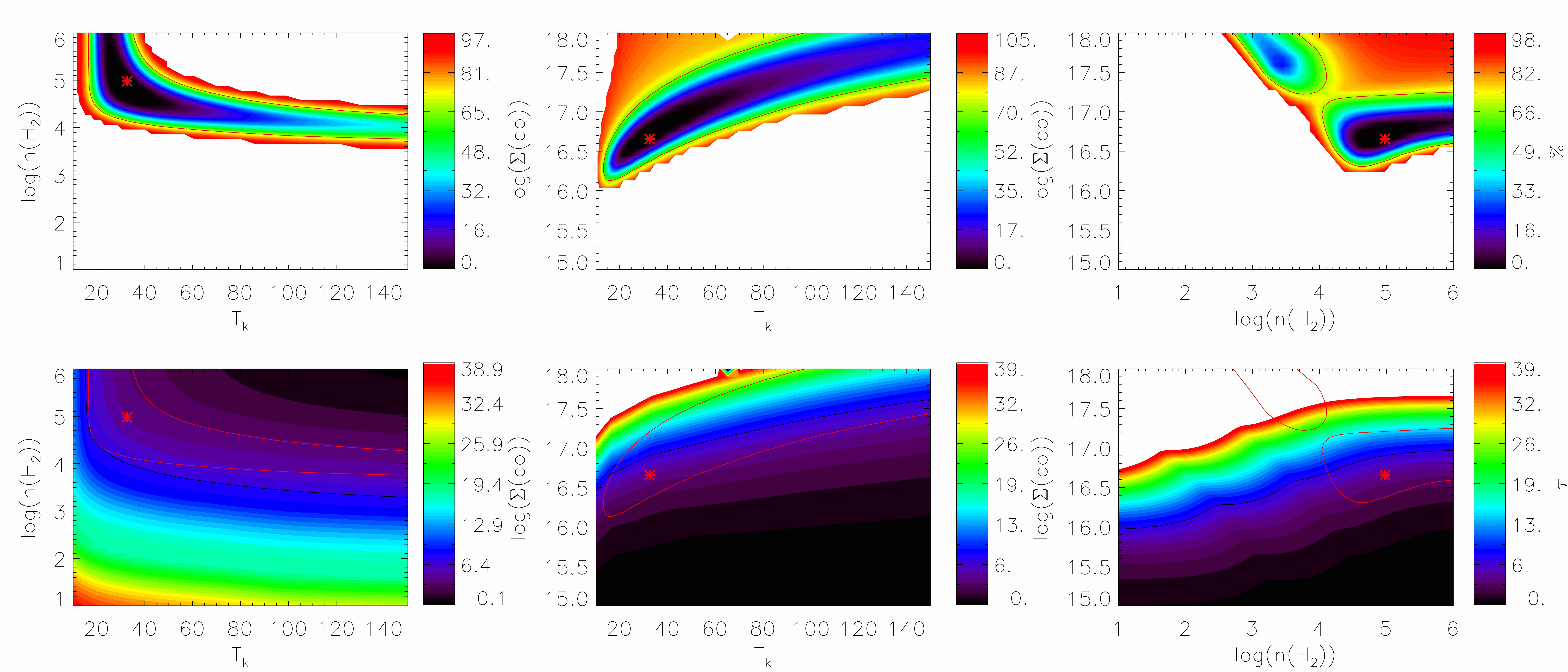}
\caption{The graphs on the top present the confidence levels of the $\chi^2$, assuming normally distributed data, around the parameters with minimum $\chi^2$. The minimum value is illustrated by the red star. The red curve marks the position of the 1$\sigma$ $\chi^2$ confidence level. The graphs on the bottom show the upper limit optical depth distribution of the lines modelled by \textsc{Radex}, see section~\ref{subsec:LVG} for details. The red curve marks  where the $\chi^2$ confidence level is equal to 1$\sigma$, and the black curve shows the $\tau=8$ contour. The best physical parameters are: T$_{\rm k}=33^{+100}_{-15}$~K, n(H$_2$)=10$^{5.0^{+1.0}_{-0.9}}$~cm$^{-3}$ and $\Sigma$(CO)=10$^{16.7^{+0.3}_{-0.6}}$~cm$^{-2}$.}
\label{fig:contmin_1c_c40_v1}
\end{figure*}

Applying the $\chi^2$ minimization method explained above, we generated the confidence levels of the $\chi^2$ around the optimal three physical parameters: temperature, density and column density (Figure~\ref{fig:contmin_1c_c40_v1}). We estimated a kinetic temperature of $33^{+100}_{-15}$~K, a molecular Hydrogen density of 10$^{5.0^{+1.0}_{-0.9}}$~cm$^{-3}$ and a $^{12}$CO column density of 10$^{16.7^{+0.3}_{-0.6}}$~cm$^{-2}$, imposing an upper limit for the optical depth of $\tau=4.9$ (in agreement with \citealt{petitpas2003}), using an abundance ratio [$^{12}$CO]/[$^{13}$CO] of 40 and an internal line width of 1~\kms. Upper limit of the optical depth means the biggest value estimated by \textsc{Radex} for all transitions. The temperature changes by 5 degrees and the molecular Hydrogen density by less that a factor 2, when changing the \mbox{[$^{12}$CO]/[$^{13}$CO]} abundance between 30 and 50.
These estimations are more robust, fully consistent with the limits presented by \citet{petitpas2003}, and present an average density larger than 10$^4$~cm$^{-3}$ as expected due from observation of HCN(1-0) in the nucleus by \citep{Kohno2003}. We estimated the errors in the magnitudes looking at 1$\sigma$ confidence level and an upper limit of the optical depth $\tau<8$.

Moreover, we obtained a $^{12}$CO column density of $4 \times 10^{16} \, \rm cm^{-2}$ for a line width of 1~\kms. If we assume a typical line width of 30~\kms \citep{Kohno1999} and a \mbox{[H$_2$]/[$^{12}$CO]} concentration ratio of \mbox{10$^4$} \citep{Toolsofradioastronomy}, we obtain a H$_2$ column density of $1.2 \times 10^{22} \, \rm cm^{-2}$, which is also in agreement with the estimation made using the \xco\ conversion factor presented in section~\ref{subsec:xco}. The kinetic temperature is in concordance with the value derived using the LTE approximation, but its uncertainty of 100~K shows a possible contribution of a warmer component.

\subsection{Inner 21\arcsec: Two-component Model}
\label{subsec:Gasgloblal2c}
The central 21\arcsec ($\sim$1~kpc) of the galaxy hosts a multitude of  molecular clouds with sizes of the order of tens of parsecs and very likely exhibiting differing physical conditions. Thus the one-component model is not realistic to describe the entire region covered by the unresolved central beam of NGC\,1097. We take a step further in an attempt to describe the typical physical properties of the cold interstellar gas using more than one component \citep[cf., ][]{Israel2009}. We presumed the inner kpc be formed by a cold (30K) and a warm (90K) component, and made a similar $\chi^2$ minimization study explained previously in this section, varying the $\rm H_2$ densities, $^{12}$CO column densities and the fraction between the two components  \mbox{$\mathcal{F}= \rm[cold \, component]/[warm + cold \, component]$}. We changed the input variables within the ranges: log$_{10}$(n(H$_2$)$_{30\rm K}$)=[3,6] (cm$^{-3}$, 20 steps), log$_{10}$(n(H$_2$)$_{90 \rm K}$)=[1.5,4.5] (cm$^{-3}$, 20 steps), log$_{10}$($\Sigma$($^{12}$CO))=[15,18] (cm$^{-2}$, 14 steps) and $\mathcal{F}$=[0,1] (8 steps). In the selection of the ranges for the densities, we imposed the cold component to be denser than the warm component.
We fixed the abundance ratio \mbox{[$^{12}$CO]/[$^{13}$CO]} to 40 and the $^{12}$CO line width to 1 \kms, and imposed the same conditions for finding the minimum $\chi^2$ value as explained previously in this section: a low optical depth for the two components and a proper value for the $\chi^2$ of all individual pairs of line ratios. Our results are shown in  table~\ref{tab:2components} with a ratio $\mathcal{F}=0.85$. It is important to note here that the $\chi^2$ analysis has been carried out using different steps for the fitted parameters and different limits for the optical depth and $\chi^2$ of the individual ratios, in order to constrain the confidence of the method. The results are correlated within the errors, making stronger their confidence.

\begin{table}
  \center
\caption{Kinetic temperature T$_{\rm k}$, column density n(H$_2$), surface density $\Sigma$($^{12}$CO) and optical depth $\tau$ from the one- and two-component models.}
\label{tab:2components}
   \centering\tabcolsep=9pt
   \begin{tabular}{ c | c | c c }
   \toprule
   & \multirow{2}{*}{\textbf{1 component}} & \multicolumn{2}{c}{\textbf{2 components}}\\
   & & cold & warm \\
   \midrule
  T$_{\rm k}$ (K) & 33$\pm^{100}_{15}$& 30 & 90\\[2mm]
  log$_{10}$(n(H$_2$)) (cm$^{-3}$) & 5.0$\pm^{1.0}_{0.9}$ & 3.9$\pm^{2.1}_{0.5}$ & 3.6$\pm^{1.0}_{0.6}$\\[2mm]
  log$_{10}$($\Sigma$($^{12}$CO)) (cm$^{-2}$) &16.7$\pm^{0.3}_{0.6}$ & 16.7$\pm^{0.3}_{0.3}$&17.1$\pm^{0.7}_{2.0}$\\[2mm]
  $\tau$($^{12}$CO) & $<4.9$ & $<6.6$ & $<7.3$ \\
   \bottomrule
\end{tabular}
\end{table}

\section{Emission lines at different velocity regions}
\label{sec:gasring}
\subsection{Dissecting the Inner 21\arcsec}
\label{subsec:gasring}
\begin{figure*}
   \centering
\includegraphics[width=0.49\textwidth, trim=5mm 8mm 40mm 0mm]{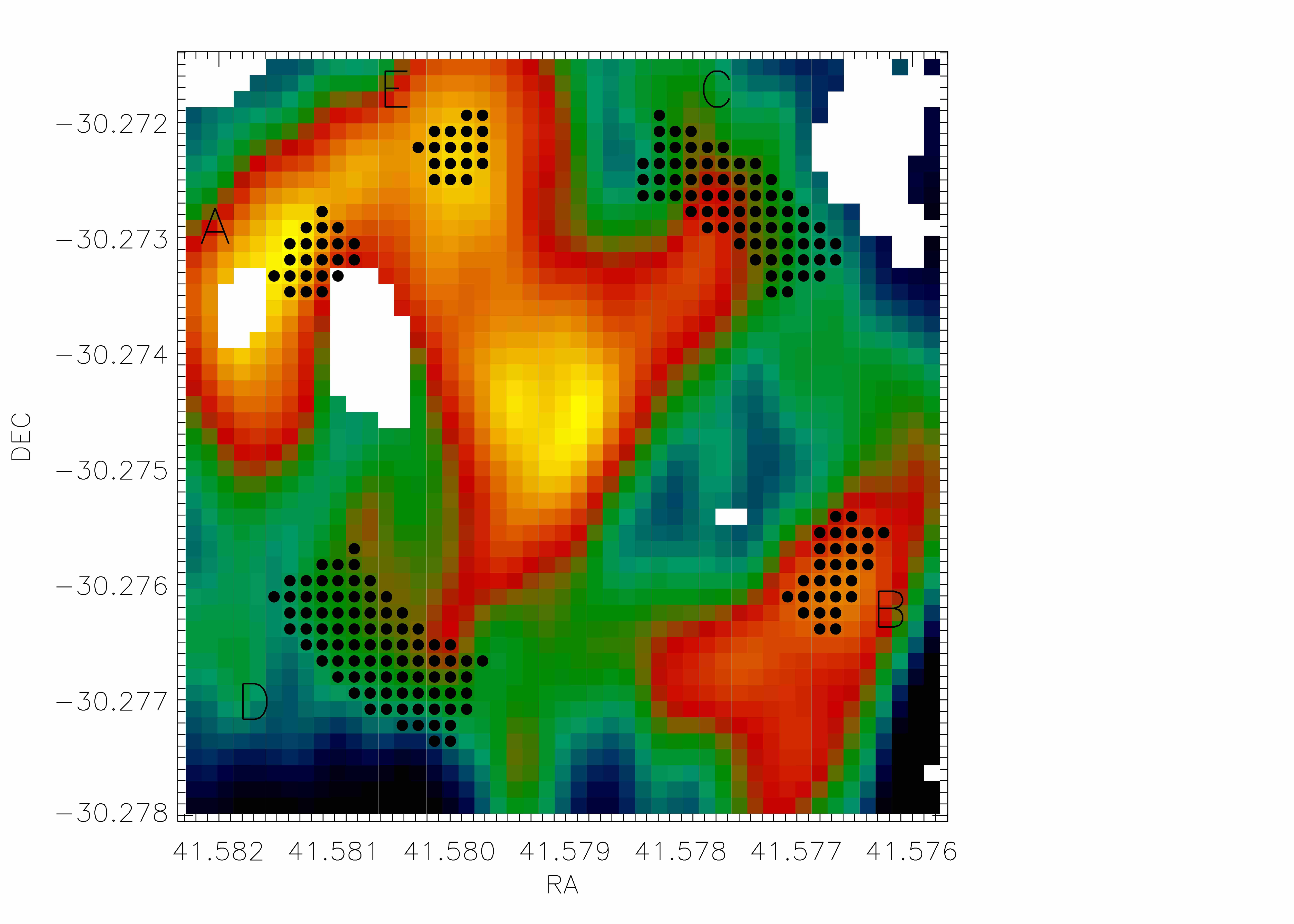}
\includegraphics[width=0.49\textwidth, trim=5mm 8mm 40mm 0mm]{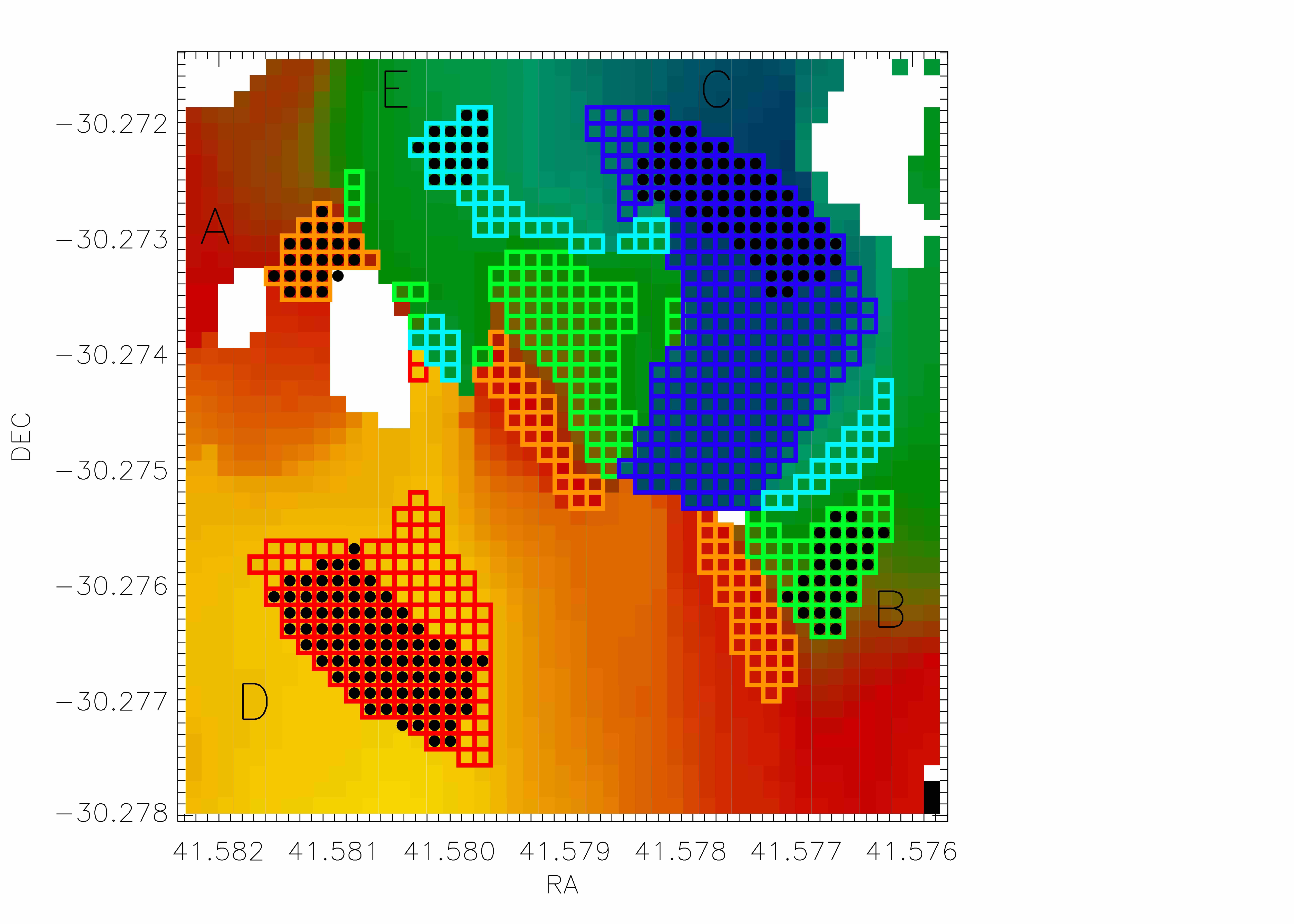}
\caption{The five regions along the inner starburst ring of NGC\,1097 with the two bar-ring connection points marked as A and B. All regions are 
over-plotted on the underlying $^{12}$CO(1-0) intensity map {\em (Left)} . All regions were selected taking into account their distinct velocities and the $S/N$ criteria as described in section~\ref{subsec:gasring}, and the full velocity window which contributes to the signal in corresponding channel is marked in the $^{12}$CO(1-0) velocity field {\em (Right)}. In the velocity field, each group of coloured squares indicate pixels with velocity inside the velocity range of each region.}
\label{fig:regionsI}
\end{figure*}

In this section, we use an innovative approach where we ``de-confuse" the single-dish data to be able to study ``resolved" emission lines for which no interferometric data have been yet obtained. 
We used the resolved interferometric intensity and kinematic maps from \citet{Kohno2003} to select five different regions along the starburst ring for a deeper analysis of their physical properties. 
Following the combination of two independent criteria, we selected different knots along the starburst ring to analyse the state of the molecular gas. The regions were selected to have a $^{12}$CO(1-0) $S/N \ge 150$ and span a range of velocity greater than one frequency/velocity channel. These choices have been made to assure that the region can be separated from other velocity channels in the unresolved single dish data, and have a $S/N$ high enough to return reliable measurements for the different transitions . Furthermore, all regions have mean line of sight velocities such that they can be separated from each other in the velocity channels of the single dish data. Two regions are located on the intersection of the bar dust lanes with the ring (regions A and B), other two regions at 90 degrees distance (in the upper and lower side of the ring, regions C and D) and a fifth region E in another maximum CO(1-0) emission between A and C and close to a connection point (see Fig.~\ref{fig:regionsI}).

\begin{figure}
   \centering
\includegraphics[width=0.47\textwidth]{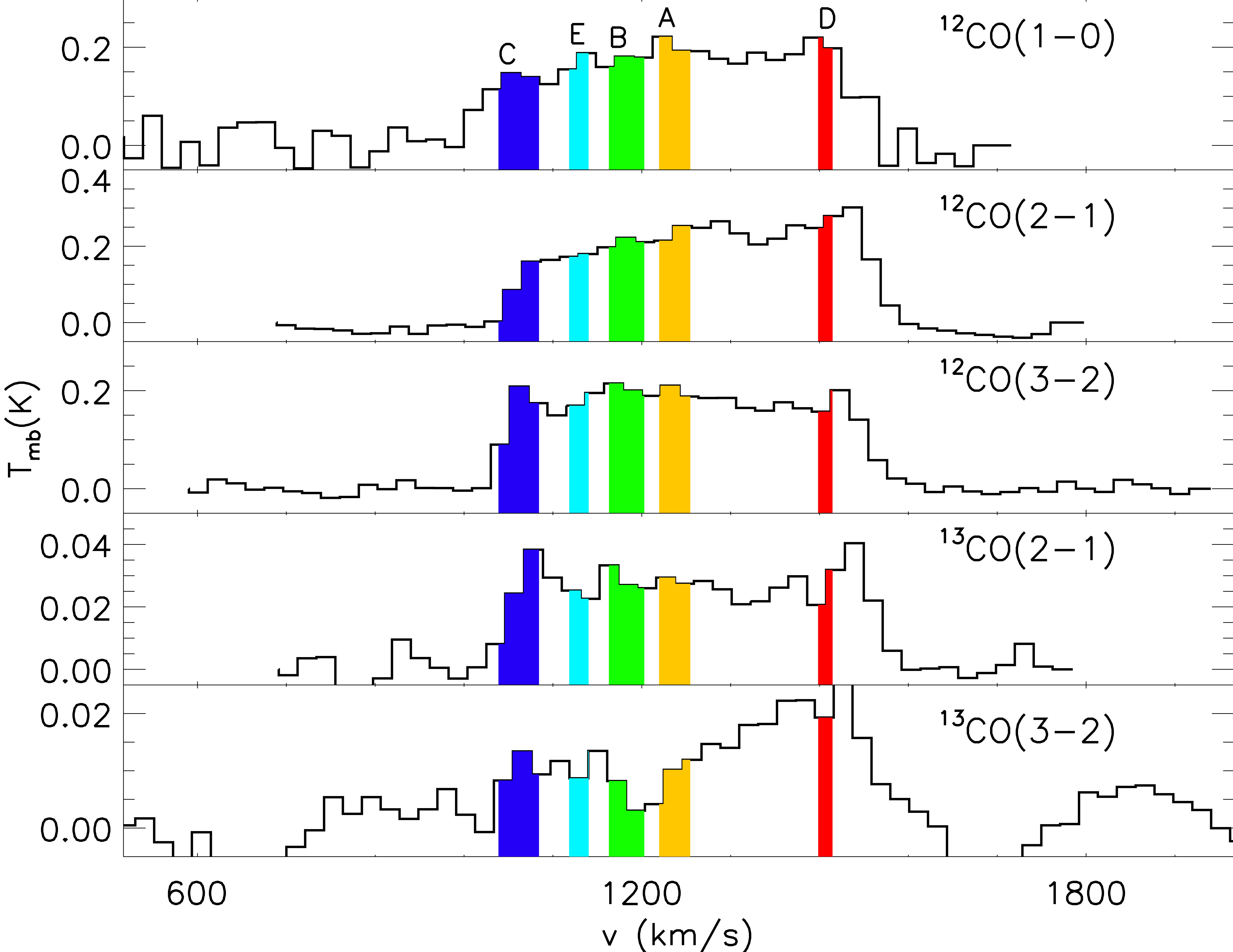}
\caption{The emission lines analysed in this paper (new APEX data, \citealt{Kohno2003} and \citealt{petitpas2003}), showing the contribution to the lines of the different regions.}
\label{fig:regionsII}
\end{figure}

To avoid unnecessary complications regarding the exact shape of each star-forming knot, we opted to select a rectangular shape along an outer radius of 10.5\arcsec\ and an inner radius between 7-8\arcsec, given their full area fits inside our beam. 
A narrow velocity range is essential for us since we aim to examine the contribution of this selected region to the global line, and it is imperative that the different regions have different velocities to distinguish them in the single dish lines. 

Once, the regions were selected and their velocity ranges were established, we studied their contribution to the global unresolved single dish emission line, see Fig.~\ref{fig:regionsII}. We took the channels which corresponded to the velocity range of each region and generated the line ratios using only these channels. We estimated the ratio's errors by adding the standard deviation of the channels to the errors listed in section~\ref{sec:obs}. With this, we have introduced the error due to a possible bad selection of the velocity intervals.
Thus, for each region we obtained four different line ratios, see table~\ref{tab:bumpsprop}. Looking at the ratios, it is possible to infer that the physical properties should be similar for every region, since the ratios are somewhat similar within the errors, but for the case of \cot/\tcot. 
To ensure the physical properties along the ring, we analysed the ratios for each region, following the procedure described in section~\ref{subsec:LVG}. We realized the \chisq\ minimization study described above for every region (see section~\ref{subsec:LVG}), imposing again that the optical depth should not exceed 5. The results are shown in the table~\ref{tab:bumpsprop}. 

Moreover, we note that the results presented in table~\ref{tab:bumpsprop} are not uniquely representative of the physical conditions of the regions from the starforming ring only. To estimate the contribution from other parts 
 which exhibit velocities which are in the range selected for each of the knots along the ring, we calculate the ratio of the $^{12}$CO(1-0) intensities from the full region inside the velocity range, and that from the ring. We find that the contribution of the signal from the ring is at best 70\% of the pixels inside the chosen velocity range (though could be as low as 35\%). This fact confirm that single dish data cannot be used to rigorously derive the physical properties of the ring only, but instead that of regions inside iso-velocity contours. Nevertheless, one can assume that the region inside the ring (i.e., galactocentric radius $<8.5$\arcsec) is much more homogeneous than then ring itself, and the fact that we have derived comparable physical properties are indicative of a rather homogeneous ring in NGC\,1097.

\renewcommand{\arraystretch}{1.8}
\begin{table*}
  \center
\caption{Properties of the different regions inside the central 21\arcsec\ region of NGC\,1097. 
}
\label{tab:bumpsprop}
   \centering\tabcolsep=20pt
   \begin{tabular}{c | c c c c c }
   \toprule
   & A & B & C & D & E\\
   \midrule
\gray  $S/N$ & 151 & 152 & 151 & 152 &150\\
  $\Delta$v (\kms) & 41 & 46 & 57 & 19 & 26 \\
\hline
\gray $\rm{\frac{^{12}CO (1-0)}{^{12}CO (2-1)}}$  & 0.87$\pm$0.19 & 0.83$\pm$0.17 & 1.3$\pm$0.7 & 0.77$\pm$0.16 & 1.0$\pm$0.2\\
$\rm{\frac{^{12}CO (3-2)}{^{12}CO (2-1)}}$ &  0.86$\pm$0.19 & 1.0$\pm$0.2 & 1.6$\pm$0.9 & 0.61$\pm$0.14 & 1.0$\pm$0.2 \\
\gray $\rm{\frac{^{12}CO (2-1)}{^{13}CO (2-1)}}$ & 8.7$\pm$1.8 & 7.5$\pm$1.7 & 4$\pm$2 & 10$\pm$1.9& 7$\pm$4 \\
$\rm{\frac{^{12}CO (3-2)}{^{13}CO (3-2)}}$ & 21$\pm$7 & 37$\pm$18 & 15$\pm$6 & 8.5$\pm$1.8 & 20$\pm$4\\
\hline
\gray  T$_k$ (K) & 17$\pm^{10}_{7}$&21$\pm^{10}_{10}$ & 12$\pm^{5}_{5}$& 21$\pm^{10}_{10}$&16$\pm^{10}_{10}$\\
  log$_{10}$(n(H$_2$)) (cm$^{-3}$) &5.0$\pm^{1.0}_{0.6}$ &5.4$\pm^{0.6}_{1.2}$ &6.0$\pm^{1.0}_{1.0}$ & 5.6$\pm^{0.8}_{0.8}$&5.0$\pm^{1.0}_{0.7}$\\
\gray  log$_{10}$($\Sigma$($^{12}$CO)) (cm$^{-2}$) &16.4$\pm^{0.3}_{0.2}$ & 16.6$\pm^{0.1}_{0.4}$&16.3$\pm^{0.3}_{0.3}$ & 16.6$\pm^{0.1}_{0.3}$&16.4$\pm^{0.3}_{0.2}$\\
   \bottomrule
\end{tabular}
\end{table*}

\subsection{The Star Forming Knots Along the Ring}
\label{sec:ring}
In star forming rings, stars are expected to form from the accreted material because of the shocks and compressions during the accretion process \citep[e.g., ][]{Elmegreen1994}, and these processes could arise randomly by gravitational instabilities without any specific order along the ring. In the case of NGC\,1097, where the ring is postulated to be at the location of the inner Lindblad resonance, such a scenario would naturally imply that the bulk of the compressive forces act on the regions where the ring meets the bar dust lanes, and thus the bar may well trigger the star formation in the ring (cf. \citealt{Knapen2006}, \citealt{Fathi2008}) at the connection points and the star formation would thereafter successively migrate along the ring following the rotation period of  $\approx18$ Myr \citep[cf., ][]{Boker2008}. The \cof\ and \cos\ emission map of the ring (\citealt{Kohno2003}, \citealt{Hsieh2008}), present clear peaks of intensity in the regions around the dust lanes connection points with the ring (A, B and E in Fig.~\ref{fig:regionsI}), and hence, may indicate stronger star formation on those spots. This could imply an increase in the molecular gas density with respect the rest of the ring.

Our improved LVG analysis using five different regions at different velocities, does not show significant differences between the molecular gas densities, temperatures or column densities, see table~\ref{tab:bumpsprop}. It is possible that the measured similarity between the derived molecular gas densities, could be due to a bad estimation of the contribution of the ring to the observed line. However, the region inside the ring (radius $<8.5$\arcsec) shows relatively small $^{12}$CO(1-0) intensity variation as well as small variations in H$\alpha$ intensity as shown in high resolution Hubble Space Telescope images presented in \citet{Fathi2006}. It is thus likely that the star forming gas inside the ring is rather homogeneously distributed, hence, its contribution to our derived physical properties along the different regions described in section~\ref{subsec:gasring} is similar. Consequently, our derived values hint that the conditions of the molecular gas at the different regions along the ring are comparable.

These results are consistent by the recent far infrared and sub-mm studies, from the Herschel Space Observatory, of the ring in NGC~1097 which showed no azimuthal age gradient \citep{HerschelII}, accompanied by an increase of the ionized gas density in over the full ring associated with star formation activity \citep{HerschelI}. Moreover, we compare the regions selected here with the resolved Very Large Array radio continuum maps \citep[][ their Figure~14]{Beck2005} and find that all star forming knots exhibit radio emission, indicative of ongoing star formation at comparable intensities. 

The strong \cof\ and \cos\ emission observed with the interferometric data (\citealt{Kohno2003} and \citealt{Hsieh2008}) on the connecting points may be due to particularly strong shocks, as predicted by the peak of the ratio [OI]63\microm/[CII]158\microm\ \citep{HerschelI}, and not necessarily due to triggered strong star formation. 

\section{Mass Inflow Rate Onto the SMBH}
\label{sec:mir}
The fuelling of the active nucleus of NGC\,1097 has been subject to much debate recently. Modelling the spectral energy distribution, \citet{Nemmen2006} estimated a inflow mass rate of $\rm \dot{m}_o = 6.4 \times 10^{-3}\, \rm \dot{M}_{Edd}$, with the Eddington accretion rate of $\rm \dot{M}_{Edd} = 2.7$ \msunyr, for a $1.2\times 10^8\rm \,M_{\odot}$ black hole. Recently, \citet{vandeVenFathi2010} used two-dimensional spectra in the optical to kinematically derive the inflow rate from ionized gas. These authors measured the inflow velocity from a kinematic analysis and the electron density from {\sc[Sii]} observations, consistent for the most inner part with \citet{Nemmen2006}. \citet{Davies2009} estimated an upper limit 1.2 \msunyr. These studies are based on simplifications of the inflow mechanism along the nuclear spiral arms and also on indirect measurements of the inflowing material by assuming that the hot ionized gas traces the underlying cold molecular gas.

\begin{figure}
	 \centering
\includegraphics[width=0.47\textwidth]{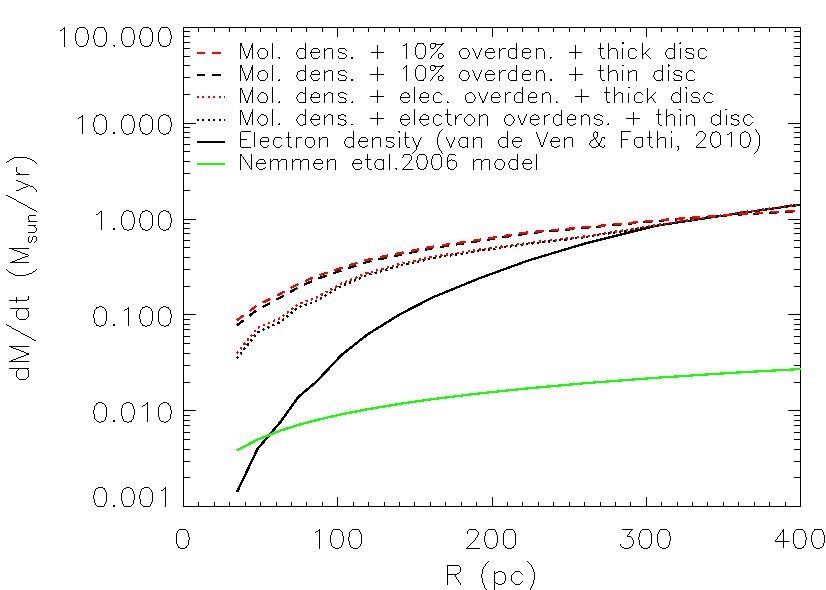}
\caption{Mass inflow rate of the inner 400 parsec radius of NGC 1097. 
\emph{Solid line} mass inflow rate estimated by \citet{vandeVenFathi2010}. 
\emph{Green line} mass inflow rate estimated by \citet{Nemmen2006}. \citet{Nemmen2006} assumed a mass inflow rate of $\rm \dot{M} = \dot{M}_o (r/r_{tr})^p$, where $p=0.8$, $r_{tr}=225$ and $\rm \dot{M}_o = 6.4\times10^{-3}\,\dot{M}_{Edd}$. 
\emph{Dotted line} our mass inflow rate estimation from the total molecular gas and using the measured of the electron over-density of the arms by \citet{vandeVenFathi2010}. 
\emph{Dashed line} our mass inflow rate estimation from the total molecular gas and using an over-density of the arms of the 10\%. 
\emph{Red dashed line} our mass inflow rate estimation from the total molecular gas and using an over-density of the arms of the 10\% and using a scale height from \citet{Romeo1994}. 
\emph{Red dotted line} our mass inflow rate estimation from the total molecular gas and using the measured of the electron over-density of the arms by \citet{vandeVenFathi2010} and using a scale height from \citet{Romeo1994}. The mass inflow rate estimated using the molecular gas density and the electron over-density for a thin disc at around 70 pc is 0.10~M$_{\odot}/$yr.}
\label{fig:massinflowrate}
\end{figure}

We estimated the mass inflow rate using our direct measurement of the molecular gas density, $\rho_{\rm gas}$, obtained for the central kpc of NGC\,1097 (see section~\ref{subsec:Gasgloblal}). We combined it with the inflow velocity estimation from \citet{vandeVenFathi2010}, and their mass inflow rate analysis.
\begin{equation}
\dot{M}=m \,  v_{\rm inflow} \Delta \rho_{\rm gas} \,\pi R^2 \,\frac{h}{R}\,\frac{1}{4m},
\label{eq:massinflowrate}
\end{equation}
where $m$ is the number of arms, $v_{\rm inflow} $ the inflow velocity, $\Delta \rho_{\rm gas}$ the over-density on the arms, $R$ the galactocentric radius and $h$ the disc scale height. We assumed a nuclear spiral arms over-density of the 10\%, i.e., $\Delta\rho_{\rm gas} = 0.1\rho_{\rm gas}$ (e.g. \citealt{EnglmaierShlosman2000}, \citealt{Maciejewski2004b}), and for comparison also measured the mass inflow rate using the electron over-density measured by \citet{vandeVenFathi2010}. We further compare the results estimating the scale height $h$ taking into account the thickness of the disc, and used the criterion of marginal stability $Q_{\mathrm{eff}}=1$ \citep{Romeo1994}, where 
\begin{equation}
Q_{\rm eff}\sim Q(1+2\zeta); \;\;\; \zeta = \frac{h \kappa^2}{2 \pi G \Sigma_{gas}}; \;\;\; Q = \frac{\kappa c_s}{\pi G \Sigma_{\rm gas}}
\end{equation}
$\kappa$ is the epicyclic frequency, $c_s$ the sound speed (we use $c_s=10$~\kms) and $\Sigma_{\rm gas} \approx \rho_{\rm gas}/h$ the gas column density. We did not find a significant difference between this assumption and the marginally stable disc, and we attribute this to the fact that we are measuring  cold molecular gas which is thinner than its ionized counterpart, i.e., the cold gas component has $h/R < 1$. We obtained several measurements for the mass inflow rate onto the SMBH in NGC\,1097 as shown in Figure~\ref{fig:massinflowrate}. 

Our measurement using the thick disc description is 0.11 M$_{\odot}/$yr at 1\arcsec= 70 pc distance from the SMBH, which is the radius to which the line of slight velocities were reliably measured by \citet{vandeVenFathi2010}. This value is still in agreement with the accretion estimation for the transition between Seyfert~1 and Low-ionization nuclear emission-line region galaxies \citep{Ho2005}, however, puts NGC\,1097 on the more efficient end of the distribution of the transition objects. Our inflow rate at 70 pc is within the upper limit derived by \citet{Davies2009} and is one order of magnitude greater than the value derived by \citet{vandeVenFathi2010}.The difference with the latter work is driven by the fact that while our cold disc remains thin ($h/R <1$), the ionized disc of \citet{vandeVenFathi2010} thickens and to avoid, this these authors applied an upper limit for the $h/R$, leading to a lowering of the mass inflow rate (c.f. equation~\ref{eq:massinflowrate}).

\section{\xco\ Conversion Factor}
\label{sec:xco}

With the analysis of marginal stability of the disc presented in section~\ref{sec:mir}, we have been able to estimate from the kinematics analysed by \citet{vandeVenFathi2010}, the column density of the gas along the nucleus using \citet{Toomre1964} (cf. \citealt{vandeVenFathi2010}) and \citet{Romeo1994} criteria. This method for estimating the surface density is completely independent of the CO emission in the galaxy. 
Assuming Q=1 (or Q$_{eff}$=1) and given $c_s$=10~\kms\ as well as $\kappa$ (from the ``kinematic model" of \citet{vandeVenFathi2010}), yields a prediction for $\Sigma_{gas}$; then assuming $\Sigma_{gas}=\Sigma_{H_2}$, the ratio with the CO intensity yields the \xco\ factor.
Hence, using this calculated column density for different radii, we could in principle study the \xco\ conversion factor for a given \cof\ intensity measured with the corresponding beam size. We translate this surface column density to molecular Hydrogen column density, dividing by a factor 1.36 corresponding to the Helium contribution to the gas.

We have the \cof\ integrated velocity intensity for a beam size of 15\arcsec\ \citep{Kohno2003} and 21\arcsec\ (see section~\ref{sec:obs}), which correspond to radius of 7.5\arcsec and 10.5\arcsec\ , respectively. 
Hence, we calculated the average surface density determined by the marginal stability criteria, over an area similar to the beam size. Combining the surface densities and \cof\ intensities for equal areas, we estimated the \xco\ conversion factor. The results were $X_{\rm CO}=(2.8\pm0.5)\times 10^{20} \, \rm cm^{-2}\, (\rm K\, km\, s^{-1})^{-1}$ at radius of  10.5\arcsec\ and $X_{\rm CO}=(5.0\pm0.5)\times 10^{20} \, \rm cm^{-2}\, (\rm K\, km\, s^{-1})^{-1}$ at radius of 7.5\arcsec. For the estimation of the \xco\ error, we took into account the contribution to the uncertainty of the epicyclic frequency $\kappa$ (uncertainty of 7\%), and of the \cof\ integrated velocity intensity for each beam (3\% for a beam size of 15\arcsec\ and 10\% for 21\arcsec). Moreover, for these calculations we assumed a sound speed equal to 10 \kms, which is a lower limit for the sound speed consider by \citet{Davies2009}, 20-40 \kms, what makes our \xco, also a lower limit. 

We obtained a \xco\ conversion factor comparable to that presented in section~\ref{subsec:xco}. Moreover, the surface density estimated by the marginal stability analysis is $(2.4\pm0.2) \times 10^{22}\, \rm cm^{-2}$, which again is in agreement with our estimations in section~\ref{subsec:xco} and section~\ref{subsec:Gasgloblal}. While our derived \xco\ conversion factor compares well with the typical values assumed in the literature, it is in disagreement with other estimations of this factor in centre of galaxies (e.g., \citealt{Strong2004}, \citealt{Israel2009}, and references therein), which are one order of magnitude lower. 
This disagreement could be due to the fact that we have not accounted for the ionized and atomic gas column density in the total gas column density estimation.

\section{Conclusions}
\label{sec:conclusions}
We have analysed single dish observations of five different transitions of the CO molecule in the central 21\arcsec\ (kpc) region of NGC\,1097. We have used \xco\ conversion factor, LTE and LVG approximation to parametrize the cold ( $\approx 30$ K) phase of the interstellar gas from which we have identified and characterized the typical physical properties inside this region. All methods have resulted in comparable values, and in agreement with other estimations made by \citet{petitpas2003} and \citet{Hsieh2008}, who used a different set of line ratios and a different LVG approximation code. We obtained a typical kinetic temperature of about 33~K, a molecular Hydrogen density of $4.9\times 10^{3}$ \msunpcthree and a CO column density of $1.2 \times 10^{-2}$ \msunpctwo. 

We have further applied a scheme which involves the assumption that circumnuclear kpc region of NGC\,1097 consists of two dominant components, a cold and a warm ($\approx 90$~K) component (cf. \citealt{Israel2009}). The two component analysis yields a 85\% cold/total ratio, where the cold gas is denser than the warm component. This ratio and the physical properties of the cold and warm gas are similar to the values found in NGC\,1068 \citep{Israel2009}. 

To gauge the robustness of these results, we dissect the spatially unresolved observations over the central kpc of NGC\,1097 in velocity space. We apply the LVG analysis on line ratios derived from five carefully selected velocity channels and find that the physical properties remain unchanged to within the errors. These five regions were selected based on the interferometric $^{12}$CO(1--0) intensity and velocity map, and all contain relatively strong star forming knots of the star burst ring. The contributing signal from the star forming knots is at least of $S/N=150$ and the contribution of each knot to the full $^{12}$CO(1--0) within the corresponding velocity bin varies between 35 and 70\%. 
Based on the interferometric data and multi-wavelength images in the literature we can assume that the cold interstellar gas in the region interior to the ring is fairly homogeneous,
 thus find it likely that the physical parameters derived from these regions relect the inherent physical properties along the starburst ring of NGC\,1097. Using the Schmidt law, the comparable gas physical properties imply similar star formation rates along the ring. Although we do not make any strong claim about this fact, it is encouraging that they are in agreement with the recent data from the Herschel Space Observatory, supporting an star formation scenario described by \citet{Elmegreen1994}. 

The rotation curve presented in \citet{vandeVenFathi2010} which reaches approximately 400 \kms\ at 8\arcsec\ radius, yields the orbital time of 18 Myr at this radius. In the simplified case, where the ring is fed by clumps of material from the bar dust lanes, followed by episodic star formation in the ring connection points with the bar dust lanes, the star forming complexes could migrate along the ring in a very short time. In such a case the star forming knots along the ring are expected to exhibit similar properties, since they ``mix'' very efficiently. This is further supported by the fact that the knots coincide with knots of enhanced radio continuum emission \citep{Beck2005}.


Making use of our new derivations of the cold gas densities, we have been able to revise the mass inflow rate onto the SMBH in NGC\,1097. We derived an inflow rate of 1.1~M$_{\odot}/$yr for a thick disc, and conclude that accounting for the total interstellar medium and applying a careful contribution of the disc thickness and corresponding stability criterion increases the inflow rate by a factor 10. The critical value \mbox{0.01~$\dot{M}_{\rm Edd}$} for the transition between LINER and Sy1 galaxies was found by \citet{Ho2005} with a distribution over three orders of magnitudes. Our new measurement of the accretion rate onto the SMBH of NGC\,1097 places this galaxy at the higher end of this distribution, and combining our derived accretion rate with the updated Eddington ratio presented in \citet{Eracleous2010}, NGC\,1097 is thus placed in the regime between the advection-dominated mass accretion and accretion via a thin disc structure (e.g., \citealt{Narayan1998}).

We have moreover calculated the \xco\ conversion factor for the centre of NGC\,1097 using an independent estimation of the surface density to the CO emission. We used the same analysis described in \citet{vandeVenFathi2010} from the kinematics of NGC\,1097. We obtained $X_{\rm CO}=(2.8\pm0.5)\times 10^{20} \, \rm cm^{-2}\, (\rm K\, km\, s^{-1})^{-1}$ at radius 10.5\arcsec\ and $X_{\rm CO}=(5.0\pm0.5)\times 10^{20} \, \rm cm^{-2}\, (\rm K\, km\, s^{-1})^{-1}$ at radius 7.5\arcsec. 
The surface density given by this ``kinematical method" and the \xco\ conversion factor are in agreement with our estimations, see section~\ref{subsec:xco} and section~\ref{subsec:Gasgloblal}.
However, these values found at the centre of NGC\,1097 are in disagreement with values found for centres of other active galaxies which are one order of magnitude lower (e.g. \citealt{Strong2004}, \citealt{Israel2009}). 

With the approach and analysis described in this paper we have demonstrated that important physical properties can be derived to a resolution beyond the single dish resolution element. To obtain a more accurate description of the molecular gas physical properties of the nuclear region of NGC\,1097, observations of higher transitions, e.g., \mbox{CO~(4--3)}, with the Atacama Large Millimetre Array (ALMA) will help overcome most of the limitations that we have faced in the current work. The ALMA data will ($i$) give us the complete description of the properties of star forming knots in the ring, ($ii$) will allow us to calculate the exact mass inflow rate for different radii by resolving the density for different radii, and ($iii$) will resolve the unknown nature of the \xco\ conversion factor in the circumnuclear kpc of this prototype LINER/Sy1 Galaxy.

\vspace{5mm}

We thank Kotaro Kohno for sending us his interferometric data, and we are indebted to Alessandro Romeo for sharing his insightful ideas on marginally stability. NP-F acknowledges financial support from NOTSA, and KF is supported by the Swedish Research Council (Vetenskapsr\aa det). This work was partly based on APEX observations (E-081.B-0298A-2008, PI: Fathi), (O-083.F-9310A-2009, PI: Fathi) and (O-086.F-9319A-2010, PI: Pi\~nol-Ferrer). Finally we thank the anonymous referee for the positive and insightful comments.

\bibliographystyle{mn2e}
\bibliography{mybib}


\end{document}